\documentstyle[12pt]{article}
\begin{document}
\title{ Berezin integration on noncompact supermanifolds \\
       }

\author{ J. Alfaro$^{\dagger}$ and  L.F. Urrutia$^{\ddagger}$,\\
$^{\dagger}$ Facultad de F\'\i sica,\\
Universidad Cat\'olica de Chile, \\
Casilla 306, Santiago 22. \\
$^{\ddagger}$Departamento de F\'\i sica de Altas Energ\'\i as,\\
Instituto de Ciencias Nucleares, \\ Universidad Nacional 
Aut\'onoma de M\'exico, \\
Apartado Postal  70-543, 04510,  M\'exico D.F.
}
\date{}

\maketitle

\begin{abstract}
 An heuristic derivation of the tranformation law for the Berezin integration
 measure
 in noncompact supermanifolds, obtained by Roshstein \cite{Ro}, is presented.
\end{abstract}

\maketitle
\baselineskip=20pt

Superanalysis is playing an increasingly important role 
in many physical applications
like supermatrix models \cite{Plefka}, conformal field theory and 
two-dimensional gravity \cite {Abdalla}, disordered system and quantum chaos
\cite{Efetov}, for example.

A central idea in superanalysis is the definition  of integration over 
a supermanifold, for which we  take the Berezin integration \cite{Ber}.
Other approaches to integration over supermanifolds are found in Refs. 
\cite{Rabin, Rogers}.

It is well known that the Berezin integration
is well defined on superfunctions which have compact support \cite{TH}.
Let us consider a coordinate system $y^i, \eta^A, i=1, \dots n, \quad A=1,
\dots, m$, where an arbitrary superfunction $F(y, \eta)$ can be written as
\begin{equation}
F(y, \eta)=\sum_{A_1, \dots A_q} F_{A_1, \dots A_q} (y^1, \dots, y^n)
\eta^{A_1} \dots \eta^{A_q}.
\end{equation}
$F$ is of compact support if all the functions 
$F_{A_1, \dots A_q} (y^1, \dots, y^n)$ vanish whenever any coordinate $y^i$ 
is on the boundary of the bosonic integration region. In this case, the 
standard
Berezin integration measure 
\begin{equation}
dy^1\dots dy^n d\eta^1\dots 
d\eta^m \equiv dy d\eta,
\label{meas}
\end{equation}
 behaves well under an arbitrary change of coordinates
$( y, \eta) \rightarrow (x, \theta)$ having the following transformation law
\begin{equation}
dy^1\dots dy^n d\eta^1\dots d\eta^m= sdet \left( {\partial(y, \eta) \over 
\partial(x, \theta)} \right)dx^1\dots dx^n d\theta^1\dots d\theta^m,
\label{scc}
\end{equation}
where $sdet \left({\partial(y, \eta) \over 
\partial(x, \theta)}\right)$ is often called the Berezinian. We are using 
the standard integration rules for the anticommuting coordinates
\begin{equation}
\int d\eta^A=0, \quad \int d\eta^A \eta^B= \delta^{AB}.
\end{equation}

When the function to be integrated is not of compact support, then the
measure (\ref{meas}) does not follow the rule (\ref{scc}) under a change of
integration variables. This can be readily
verified in the well-known example $ F=y, \quad 0<y<1$, 
under the change of variables 
$y=x + \theta^1\theta^2, \quad \eta^1=\theta^1,\quad
\eta^2=\theta^2$  \cite{Ro, TH}.  This problem has already been adressed by 
Berezin, 
among others. In Ref. \cite{Ber}, he provides an explicit formula for the
transformation rule in this case. An alternative formulation
is presented in Ref. \cite{Ro}, 
together with an extension of the Berezin integration.

In this letter we  rederive the transformation law discovered in 
Ref\cite{Ro} in an  
heuristical way. To this end,  let us
introduce a regulator function $R_\lambda(y^1, \dots, y^n)$ such that
\begin{equation}
lim_{\lambda\rightarrow \lambda_0}R_\lambda(y^1, \dots, y^n)=1, 
\quad {\rm for \ all
} \  y^i
\label{reg1}
\end{equation}
and having the property that the product
\begin{equation}
R_\lambda(y^1, \dots, y^n) F(y^1, \dots, y^n, \eta^1,\dots, \eta^m)
\label{cs1}
\end{equation}
is of compact support. 
An explicit example of the regulator defined above, for adequately 
behaved
functions $F$ of noncompact support and with $\lambda_0\rightarrow 
\infty$,  is
\begin{equation}
R_\lambda(y^1, \dots, y^n)=\Pi_{i=1}^n \left(1-{\rm e}^
{-\lambda(b^i- y^i)}\right)
\left(1-\rm{e}^{-\lambda(y^i- a^i)}\right),
\end{equation}
where $ a^i < y^i < b^i$ define the bosonic integration region.

Let us start then with the identity
\begin{eqnarray}
&&\int dy d\eta F(y, \eta)= lim_{\lambda\rightarrow\lambda_0}
\int dy d\eta R_\lambda(y) F(y,\eta)  \nonumber  \\  
&=& lim_{\lambda\rightarrow\lambda_0}
\int dx d\theta R_\lambda(y(x, \theta)) 
F(y( x, \theta),  \eta( x, \theta)) sdet\left({\partial(y, \eta) \over 
\partial(x, \theta)}\right)
\label{sp}  
\end{eqnarray}
valid for functions of compact support. In the following  call 
$$G(x, \theta)=F(y( x, \theta),  \eta( x, \theta)) 
sdet\left( {\partial(y, \eta) \over \partial(x, \theta)}\right).$$
  
In order to disentangle the RHS of (\ref{sp}), let us
focus upon the regulator. Following Ref. \cite{Ro}, we define the change 
of 
variables by
\begin{equation}
(y, \eta) = {\rm e}^Y (x, \theta) {\rm e}^{-Y}:= (x + \Delta x, \theta + 
\Delta
 \theta),
\label{chov}
\end{equation}
where $Y(x, \theta) $ is a differential operator in the
coordinates $x^i, \theta^A$. In this way, we obtain
\begin{eqnarray}
\label{f0}
R_\lambda(y(x, \theta))&=&R_\lambda({\rm e}^Y \ x \ {\rm e}^{-Y})=
{\rm e}^Y R_\lambda(x) {\rm e}^{-Y} \nonumber \\
&:=&\sum_{i_1,\dots i_q} g^{i_1,\dots i_q}
{ \partial^q \over \partial x^{i_1} \dots \partial x^{i_q}} R_\lambda(x)=
g^I(x, \theta) {\partial \over {\partial x}^I}R_\lambda(x), 
\end{eqnarray}
where  sum over repeated indices is understood.
We are explicitly using the notation of Ref. \cite{Ro} 
in the last equality of the above equation. On the other hand, 
$R_\lambda(y(x, \theta))=R_\lambda(x + \Delta x )$, so that the functions
$ g^I$ are nothing but the Taylor expansion coefficientes of 
$R_\lambda(x + \Delta x )$. In fact we have
\begin{equation}
g^{i_1,\dots i_q}(x, \theta) = {1\over q!} \Delta x^{i_1} \dots \Delta x^{i_q}.
\end{equation}

Then, the RHS of Eq.(\ref{sp}) reduces to
\begin{eqnarray}
&&lim_{\lambda\rightarrow\lambda_0}
\int dx d\theta R_\lambda(y(x, \theta)) G(x, \theta) \nonumber \\
&=&lim_{\lambda\rightarrow\lambda_0}
\int dx d\theta 
\left({\partial \over \partial x^I}R_\lambda(x)
\right) g^I(x, \theta) G(x, \theta),
\label{f1}
\end{eqnarray}
where the above relation is obtained upon sustitution of  the expression 
(\ref{f0}) in the previous line.
The next step is to integrate by parts Eq.(\ref{f1}) and subsequently 
take the
$\lambda \rightarrow \lambda_0$ limit. The contributions at the boundary 
vanish
in virtue of the condition (\ref{cs1}), having in mind that the Berezinian 
together
with the functions $g^I$ are regular there. Inside the integral, we can now
use the property (\ref{reg1}) of the regulator, obtaining
\begin{equation}
\int dy d\eta F(y, \eta)= \int dx d\theta \ (-1)^{|I|} {\partial 
\over {\partial x}^I} 
\left( g^I(x, \theta) \
sdet \left( {\partial(y, \eta) \over 
\partial(x, \theta)} \right) \ F(x, \theta)\right),
\label{fex1}
\end{equation} 
where $|I|=| i_1, \dots, i_q| = q $. 

Introducing the notation of Ref.\cite{Ro},
\begin{equation}
D_I(x, \theta)= (-1)^{|I|} \ dx \  d\theta \ {\partial\over {\partial x}^I},
\label{liop} 
\end{equation}
the  expression (\ref{fex1}) can be rewritten as
\begin{equation}
D_0(y, \eta) =  D_I(x, \theta) \ 
g^I(x, \theta) \
sdet\left( {\partial(y, \eta) \over 
\partial(x, \theta)} \right),
\end{equation}
where each $ D_I$ is an operator acting over all the functions to the right. 

Starting from  an  operator $D_I(y, \eta)$ and repeating the  above procedure
for the change of 
integration variables, we obtain
\begin{equation}
D_I(y, \eta) =  D_{I+J}(x, \theta) \ 
g^J(x, \theta) \
sdet\left( {\partial(y, \eta) \over 
\partial(x, \theta)}\right),
\end{equation}
which is the general result derived by Roshstein \cite{Ro}.
\vskip .3 cm
\noindent
{\bf Acknowledgements}

\vskip.3cm

LFU was supported in part by the projects CONACyT-CONICyT  
E120-1810 and
DGAPA-UNAM-IN100397. The work of JA  is partially supported by Fondecyt 
1980806
and the CONACyT-CONICyT project 1997-02-038.

\end{document}